\documentclass{Interspeech2024}
\usepackage{booktabs} 
\usepackage{multirow} 

\interspeechcameraready

\title{Analyzing Speech Unit Selection for Textless Speech-to-Speech Translation}

\name[affiliation={1}]{Jarod}{Duret}
\name[affiliation={1}]{Yannick}{Esteve}
\name[affiliation={2}]{Titouan}{Parcollet}

\address{
  $^1$LIA - Avignon Universite, France\\
  $^2$University of Cambridge, United-Kingdom}
\email{first@univ-avignon.fr}


\begin{document}

\maketitle

\begin{abstract}
    
    Recent advancements in textless speech-to-speech translation systems have been driven by the adoption of self-supervised learning techniques. 
    Although most state-of-the-art systems adopt a similar architecture to transform source language speech into sequences of discrete representations in the target language, the criteria for selecting these target speech units remains an open question.
    This work explores the selection process through a study of downstream tasks such as automatic speech recognition, speech synthesis, speaker recognition, and emotion recognition.  
    Interestingly, our findings reveal a discrepancy in the optimization of discrete speech units: units that perform well in resynthesis performance do not necessarily correlate with those that enhance translation efficacy.  
    This discrepancy underscores the nuanced complexity of target feature selection and its impact on the overall performance of speech-to-speech translation systems.  
    
\end{abstract}

\section{Introduction}
Speech-to-speech translation (S2ST) provides a powerful means of overcoming the communication gap between people speaking different languages by enabling effective communication across diverse languages and cultures.  
Several approaches have been proposed in the literature, including cascaded approaches~\cite{lavie1997, nakamura2006} that combine automatic speech recognition (ASR), machine translation (MT) and text-to-speech (TTS).
More recently, textless approach which leverages discrete speech units extracted from self-supervised representation has been introduced ~\cite{Lee2022direct, duret2023enhancing}.  
This technique is specifically designed to capture the linguistic content of the target speech effectively while minimizing the influence of the speaker's prosodic features.  
Previous studies demonstrated that the use of discrete speech units effectively separates linguistic content from prosodic characteristics and speaker identity.  
However, an open question remains regarding the construction and selection of these discrete speech units.  
In ~\cite{Lee2022direct, duret2023enhancing}, the authors opted to utilize HuBERT~\cite{Hsu2021hubert}, as this model has demonstrated superior performance in automatic speech recognition (ASR), spoken language modeling, and speech synthesis compared to other unsupervised representations, as shown in  ~\cite{yang2021superb}.  
Although this superiority has been established for continuous representations, there is a notable lack of analysis concerning discrete ones, especially on a layer-wise basis.

Furthermore, available speech translation corpora, such as Fisher and CALLHOME~\cite{Post2014} or CoVoST 2~\cite{Wang2021Covost}, do not contain parallel speech, the target language speech must be synthesized from text translations.    
A few datasets already provide TTS-synthesized speech, such as CVSS~\cite{Jia2022Cvss}, and more recently, SpeechMatrix~\cite{duquenne2022}, a large-scale multilingual corpus containing real speech.  
The common issue with all these datasets is the mismatch in speaker identity and emotion, they are not consistent across the source and target speech. 
This inconsistency underscores the need to take these elements into account when selecting the target speech representation.  

In this study, we explore the challenge of choosing effective discrete units for textless speech-to-speech translation.  
Additionally, we evaluate discrete self-supervised representations from various encoders reported in the literature across four downstream tasks: automatic speech recognition, speech synthesis, speaker recognition, and emotion recognition.  
Then, we investigate the potential of using semantically aligned (speech-text) speech representations to improve the ability of discrete speech units to preserve semantic information.  
This approach aims to enhance robustness against acoustic variations that could otherwise lead to diminished translation performance.

\section{Method}
For this study, we considered all existing models in the literature for speech-to-speech translation (S2ST) to the best of our knowledge.  
Consequently, two self-supervised encoders have been selected: Wav2vec 2.0~\cite{Conneau2020} and HuBERT~\cite{Hsu2021hubert}, each available in both monolingual and multilingual versions.  
Additionally, SAMU-XLSR~\cite{Khurana2022samu}, a distilled version of Wav2Vec XLS-R~\cite{babu2021xlsr} fine-tuned to predict text embeddings from a LaBSE~\cite{feng2022languageagnostic} text encoder, is also included in our analysis.
All considered models generate output at the same frequency, producing a representation of size D every 20 ms of the audio signal. For the Large versions, D = 1,024, and for the Base versions, D = 768.
The models are based on very similar Transformer-based architectures, yet they differ in their pretraining pretext tasks.  
The training of Wav2vec 2.0 is based on the contrastive predictive coding~\cite{oord2018representation} (CPC) objective, which aims to maximize the mutual information between a set of context features and predicted future samples.  
Meanwhile, HuBERT's approach involves mapping unlabeled audio to sequences of pseudo-labels obtained through the clustering of previous representations.
To extract the sequence of speech units, we employ k-means clustering on the raw speech features, using the learned centroids of the $K$ clusters to convert audio into a sequence of cluster indices for every 20ms segment of the input audio signal.  
For the base model, we extract representations from every second layer, and for the large model, from every fourth layer, to maintain manageable experiment scales.  
Another parameter is the choice of $k$. In line with prior research, we explore three values of $k$: 128, 512, and 1024. This approach allows us to assess the impact of cluster granularity on the performance of our downstream tasks.

\subsection{Downstream Tasks and Datasets}
To align with previous self-supervised learning (SSL) studies, we evaluate discrete speech representations across various tasks assessing different aspects of the speech signal  
We present four tasks designed to analyze aspects related to phonetics, speaker identity, emotions, and semantics. \\

\noindent
\textbf{Emotion Recognition (ER):} ESD~\cite{esd}, a multilingual emotional database, consists of 350 parallel utterances recorded by 10 native English and 10 native Chinese speakers (10 females, 10 males), containing five emotional states (neutral, happy, angry, sad, and surprise).  
In this study, we focus exclusively on the English subset. The official training, development, and testing splits are utilized for evaluation, with accuracy serving as the evaluation metric. \\

\noindent
\textbf{Automatic Speech Recognition (ASR):} LibriSpeech~\cite{panayotov2015librispeech}, a corpus of approximately 1000 hours of 16kHz read English speech derived from read audiobooks.  
In this study, we concentrate on the train-clean-100 subset for training.  
The dev-clean subset is used for validation, while the test-clean and test-other subsets are employed for testing.  
Character Error Rate (CER) serves as the error metric.  \\

\noindent
\textbf{Automatic Speaker Verification (ASV):} VoxCeleb1~\cite{nagrani2017voxceleb}, a large-scale speaker identification dataset, contains over 100,000 utterances from 1,251 celebrities, extracted from videos uploaded to YouTube.  
Official training, development, and testing splits are utilized for evaluation, ensuring no overlap between speakers in the training and testing sets.  
The evaluation metric is the Equal Error Rate (EER).  \\

\noindent
\textbf{Speech Synthesis:} LJSpeech~\cite{ljspeech17}, a dataset comprising 13,100 short audio clips from a single speaker reading passages from 7 non-fiction books, totaling approximately 24 hours.  
We randomly split the dataset into training, development, and testing sets with a ratio of 80:10:10\%.  
The evaluation metric is the Mean Opinion Score (MOS).  
Given the number of models, we opted for the UTokyo-SaruLab MOS prediction system~\cite{saeki2022utmos} to automatically assess the quality of the trained models.

\subsection{Systems description}

This section provides a brief description of the downstream probes employed in our study.  
For all downstream tasks, the discrete tokens are
initially passed through an embedding layer that is randomly initialized.  
The code for all experiments, training logs, and hyperparameters will be accessible once the review process has been completed. \\

\noindent
\textbf{ER \& ASV:} For the classification tasks, we follow previous benchmarks~\cite{zaiem2023speech, yang2021superb} and utilize ECAPA-TDNN, which combines convolutional and residual blocks. This system is trained using negative log-likelihood loss for ER and Additive Margin Softmax Loss for ASV. \\

\noindent
\textbf{ASR:} In the speech recognition task, we replicate a previously established benchmark~\cite{zaiem2023speech, yang2021superb}, utilizing a vanilla 2-layer BiLSTM with 1,024 units each, followed by a linear layer that maps audio to characters. The system is trained using the Connectionist Temporal Classification (CTC) loss at the character level. \\

\noindent
\textbf{Speech Synthesis:} Following~\cite{Polyak2021}, we use the HiFi-GAN neural vocoder \cite{Kong2020} to synthesize speech.
HiFiGAN is a generative adversarial network (GAN) consisting of one generator and a set of discriminators.
We adapted the generator architecture to take as input a sequence of discrete-unit. \\

\subsection{Speech to Speech Translation}
The following section describes the dataset and the speech-to-unit translation (S2UT) model used to assess the choice of discrete speech units.  
We use the CVSS corpora to train and evaluate our speech-to-unit translation model.
CVSS is a massively multilingual-to-English speech-to-speech translation corpus, covering pairs from 21 languages to English.  
However, only the French-to-English translation is considered in this study.  
The dataset includes two versions of spoken translation: CVSS-C and CVSS-T. While both versions can be utilized to train our system, we use CVSS-C because of its superior speech quality.  
Official training, development and testing splits are utilized for evaluation.  
We build the S2UT model by adapting the transformer encoder-decoder framework presented in~\cite{Popuri2022}.
The encoder is composed of a Wav2Vec 2.0 base pre-trained on 3K hours of French speech~\footnote{huggingface.co/LeBenchmark/wav2vec2-FR-3K-base}.
As a decoder, we use $6$ transformer layers with a random weight initialization. 

We combined the Wav2Vec 2.0 encoder along with the transformer decoder and we finetune the whole model end-to-end.  
During inference, the S2UT model's predictions are fed into a vocoder trained on discrete speech units for speech synthesis
Recent research in speech-to-speech translation advocates for using BLEU scores to evaluate translation quality.
First, we use a speech recognition model~\footnote{huggingface.co/speechbrain/asr-transformer-librispeech} to compute the transcriptions of the generated speech.  
Then, we compute the BLEU score for the ASR-decoded text in comparison to the reference translations.  
We acknowledge that the ASR BLEU score may be influenced by ASR model performance.

\section{Results and Discussion}
In the following section, we first discuss the results of the four downstream tasks independently.
Next, we evaluate the translation quality of the retained encoders and k-means (k=number of discrete speech units) against a baseline setup reproduced from the literature.  
Finally, we discuss the correlation between downstream task and speech-to-speech translation performance.  
In the following tables, for Base model, we report scores from every second layer, and for Large model, from every fourth layer.

\subsection{Emotion recognition}
From Table~\ref{tab:er}, we can observe similar performance for HuBERT and Wav2Vec2.  
Both encoders start with relatively high accuracy in initial layers, indicating their capability to capture emotional content effectively at these stages.  
The best performance is observed with the HuBERT Base model using k=1024 and layer~$2$, achieving the highest accuracy of $66.1\%$.  
We denote a progressive decrease in performance as layers progress, this decrease is especially pronounced in the Wav2Vec2 Base model, where accuracy drops significantly from initial to subsequent layers, highlighting a potential issue in maintaining emotional content representation in deeper layers.
For Multilingual encoders, Wav2Vec2 XLS-R generally achieves better performance across most layers.  
The decline in SAMU-XLSR performance across consecutive layers likely stems from its specialization in encoding semantic information more effectively in the upper layers, albeit at the cost of less efficient encoding of emotional information.
In addition, due to the presence of identical linguistic in the utterances for all emotional states in the dataset, the model struggles to effectively utilize semantic information for accurate label prediction.

\begin{table}[!ht]
 \centering
 \caption{Benchmarking results for the emotion recognition task across various self-supervised learning (SSL) models, both in base and large configurations, using different cluster sizes (k=128, 512, 1024) for speech unit extraction. The Accuracy is used as the performance metric}
 \label{tab:er}
 \resizebox{0.5\textwidth}{!}{
 \begin{tabular}{c|c|cccccc}
    \toprule
    \multirow{2}{*}{SSL Model} & \multirow{2}{*}{Setting} & \multicolumn{6}{c}{Layer Base/Large - ACC $\uparrow$} \\
     & & 2/4 & 4/8 & 6/12 & 8/16 & 10/20 & 12/24 \\
     \midrule
     \multirow{3}{*}{Hubert Base} & k=128 & 64.4 & 62.1 & 60.8 & 62.7 & 54.6 & 60.1 \\
     & k=512 & 65.4 & 64.2 & 59.5 & 56.9 & 57.3 & 63.7 \\
     & k=1024 & \textbf{66.1} & 63.3 & 62.7 & 55.3 & 59.9 & 63.0 \\
     \midrule
     \multirow{3}{*}{Wav2Vec2 Base} & k=128 & 63.8 & 54.8 & 49.7 & 48.3 & 44.5 & 38.8 \\
     & k=512 & 62.8 & 56.9 & 49.1 & 47.4 & 44.7 & 36.1 \\
     & k=1024 & 65.9 & 59.6 & 48.5 & 43.8 & 42.2 & 36.0  \\
     \midrule
     \midrule
     \multirow{3}{*}{mHuBERT Base} & k=128 & 65.2 & 61.4 & 55.7 & 52.3 & 54.4 & 57.9 \\
     & k=512 & 64.2 & 63.3 & 57.9 & 55.3 & 59.3 & 59.2 \\
     & k=1024 & 62.7 & 63.8 & 58.3 & 56.5 & 57.4 & 58.5 \\
     \midrule
     \multirow{3}{*}{Wav2Vec2 XLS-R Large} & k=128 & 65.7 & 64.5 & 65.1 & 66.9 & 69.7 & 51.3 \\
     & k=512 & 66.3 & \textbf{69.6} & 67.7 & 66.9 & 68.3 & 56.4 \\
     & k=1024 & 69.3 & 66.6 & 65.4 & 66.4 & 67.7 & 57.9  \\
     \midrule
     \multirow{3}{*}{SAMU-XLSR Large} & k=128 & 65.5 & 48.1 & 42.3 & 36.8 & 31.1 & 29.6 \\
     & k=512 & 65.4 & 48.1 & 39.3 & 34.3 & 31.3 & 30.1 \\
     & k=1024 & 65.4 & 49.1 & 40.1 & 34.3 & 30.1 & 30.5 \\
    \bottomrule
 \end{tabular}
 }
\end{table}

\subsection{Automatic speech recognition}
As expected for the Automatic Speech Recognition (ASR) task, Table~\ref{tab:asr} illustrates that discrete speech units generated by the high middle layers generally yield superior results in Wav2Vec 2.0 models. Regarding continuous speech representation, \cite{pasad2021layer} has demonstrated that linguistic word-level information is better encoded in the high middle layers of these self-supervised learning (SSL) models. HuBERT exhibits a distinct behavior, as the discrete speech units generated in its deepest layers (excluding the last one) achieve optimal performance, consistent with observations for continuous speech representations presented in~\cite{pasad2023comparative}.

Wav2Vec2 Base consistently outperforms HuBERT Base across all cluster sizes, achieving the lowest Character Error Rate (CER) of $1.38$ at layer $10$ with $k=1024$. 
The multilingual HuBERT model demonstrates a decrease in performance compared to its monolingual counterpart, indicating potential challenges in handling diverse languages for ASR tasks.
Notably, Wav2Vec2 XLS-R surpasses SAMU-XLSR across all layers, despite SAMU-XLSR being fine-tuned for text embedding prediction.
Furthermore, we observe that the impact of cluster size is more pronounced in the larger models than in the base models, suggesting that larger models benefit from a higher number of clusters.

\begin{table}[!h]
 \centering
 \caption{Benchmarking results for the automatic speech recognition task across various self-supervised learning (SSL) models, both in base and large configurations, using different cluster sizes (k=128, 512, 1024) for speech unit extraction. The Character Error Rate (CER) is used as the performance metric.}
 \label{tab:asr}
 \resizebox{0.5\textwidth}{!}{
 \begin{tabular}{c|c|cccccc}
    \toprule
    \multirow{2}{*}{SSL Model} & \multirow{2}{*}{Setting} & \multicolumn{6}{c}{Layer Base/Large - CER $\downarrow$} \\
     & & 2/4 & 4/8 & 6/12 & 8/16 & 10/20 & 12/24 \\
     \midrule
     \multirow{3}{*}{Hubert Base} & k=128 & 13.44 & 9.15 & 6.10 & 5.14 & 4.64 & 5.70 \\
     & k=512 & 11.36 & 8.34 & 5.54 & 4.03 & 3.37 & 4.82 \\
     & k=1024 & 11.44 & 8.14 & 5.50 & 3.94 & 3.16 & 4.96 \\
     \midrule
     \multirow{3}{*}{Wav2Vec2 Base} & k=128 & 10.66 & 6.54 & 4.49 & 3.96 & 1.49 & 4.31 \\
     & k=512 & 9.44 & 5.94 & 4.00 & 2.68 & 1.40 & 4.37 \\
     & k=1024 & 9.45 & 6.06 & 3.90 & 2.54 & \textbf{1.38} & 4.24  \\
     \midrule
     \midrule
     \multirow{3}{*}{mHuBERT Base} & k=128 & 15.03 & 10.22 & 7.36 & 6.79 & 6.66 & 6.04 \\
     & k=512 & 12.81 & 9.16 & 6.89 & 6.23 & 5.96 & 5.74 \\
     & k=1024 & 12.79 & 9.03 & 6.98 & 6.23 & 5.87 & 5.69 \\
     \midrule
     \multirow{3}{*}{Wav2Vec2 XLS-R Large} & k=128 & 16.33 & 11.08 & 9.15 & 5.72 & 15.98 & 45.35 \\
     & k=512 & 12.32 & 8.09 & 6.73 & 4.30 & 10.28 & 32.86 \\
     & k=1024 & 11.87 & 7.82 & 6.26 & \textbf{4.07} & 9.09 & 29.22  \\
     \midrule
     \multirow{3}{*}{SAMU-XLSR Large} & k=128 & 11.70 & 6.66 & 11.43 & 16.89 & 27.01 & 71.41 \\
     & k=512 & 10.06 & 6.28 & 6.36 & 7.80 & 13.90 & 66.91 \\
     & k=1024 & 9.89 & 6.14 & 5.72 & 5.77 & 10.13 & 70.94 \\
    \bottomrule
 \end{tabular}
 }
\end{table}

\subsection{Automatic speaker verification}
From Table~\ref{tab:asv}, we denote that HuBERT Base model systematically outperforms the Wav2Vec2 Base across different layers and cluster sizes.  
The table also illustrates a notable increase in the EER for SAMU-XLSR at higher layers across all cluster sizes, significantly underperforming compared to Wav2Vec2 XLS-R.  
This trend suggests a loss of speaker-specific information in SAMU-XLSR's higher layers, which might be due to its focus on retaining semantic information.  
We observe a consistent pattern from the data, showing that models generally achieve better performance at lower to mid layers than at the highest layers for speaker verification tasks.  
The impact of cluster size on EER varies across models, but the general improvement in performance with increasing cluster size suggests that more granular speech unit representations can enhance ASV performance.

\begin{table}[!h]
 \centering
 \caption{Benchmarking results for the automatic speaker verification task across various self-supervised learning (SSL) models, both in base and large configurations, using different cluster sizes (k=128, 512, 1024) for speech unit extraction. The Equal Error Rate (EER) is used as the performance metric.}
 \label{tab:asv}
 \resizebox{0.5\textwidth}{!}{
 \begin{tabular}{c|c|cccccc}
    \toprule
    \multirow{2}{*}{SSL Model} & \multirow{2}{*}{Setting} & \multicolumn{6}{c}{Layer Base/Large - EER $\downarrow$} \\
     & & 2/4 & 4/8 & 6/12 & 8/16 & 10/20 & 12/24 \\
     \midrule
     \multirow{3}{*}{Hubert Base} & k=128 & 19.43 & 19.88 & 21.26 & 21.25 & 20.70 & 19.05 \\
     & k=512 & 17.73 & 18.28 & 19.99 & 21.19 & 19.20 & 16.54 \\
     & k=1024 & 17.11 & 18.54 & 19.93 & 21.26 & 18.28 & \textbf{16.46} \\
     \midrule
     \multirow{3}{*}{Wav2Vec2 Base} & k=128 & 21.10 & 23.20 & 26.65 & 27.25 & 31.27 & 33.35 \\
     & k=512 & 19.58 & 22.50 & 26.42 & 26.96 & 30.63 & 33.68 \\
     & k=1024 & 19.27 & 23.15 & 26.34 & 27.17 & 30.06 & 33.03  \\
     \midrule
     \midrule
     \multirow{3}{*}{mHuBERT Base} & k=128 & 19.07 & 21.50 & 23.97 & 25.46 & 25.36 & 23.79 \\
     & k=512 & 18.49 & 20.09 & 21.83 & 23.97 & 24.66 & 22.07 \\
     & k=1024 & 17.34 & 20.27 & 23.03 & 23.93 & 23.60 & 22.74 \\
     \midrule
     \multirow{3}{*}{Wav2Vec2 XLS-R Large} & k=128 & 20.14 & 21.48 & 23.68 & 20.83 & 23.38 & 38.67 \\
     & k=512 & \textbf{18.07} & 19.49 & 21.66 & 22.49 & 22.78 & 38.58 \\
     & k=1024 & 18.79 & 20.92 & 23.59 & 19.62 & 22.51 & 37.01  \\
     \midrule
     \multirow{3}{*}{SAMU-XLSR Large} & k=128 & 19.49 & 26.21 & 31.53 & 36.11 & 41.71 & 46.11 \\
     & k=512 & 18.19 & 23.84 & 29.90 & 34.46 & 39.74 & 45.55 \\
     & k=1024 & 18.21 & 26.11 & 29.30 & 33.73 & 39.59 & 45.84 \\
    \bottomrule
 \end{tabular}
 }
\end{table}

\subsection{Speech Synthesis}
In Table~\ref{tab:ss}, we can see that the vocoder fed by the discrete speech units computed from HuBERT Base model outperforms the one fed by the discrete speech units generated by Wav2Vec2 Base model, particularly on layer~$6$ with a cluster size of $512$, achieving a MOS score of $3.45$.  
The highest score at layer $6$ aligns with previous work's on speech-to-speech translation, highlighting the significance of this configuration for achieving high-quality speech synthesis.  
Among the multilingual models, Wav2Vec2-XLS-R outperforms both mHuBERT Base and SAMU XLSR Large with the highest MOS score of $3.80$ on layer 20 with a cluster size of $512$. 
This indicates a potential benefit of larger model and extensive training data, meanwhile mHuBERT base shows competitive performance, especially with a cluster size of $512$.
Finally, SAMU-XLSR Large demonstrates a significant decline in MOS scores at higher layers and for all cluster sizes.  
This drop can be attributed to the model's focus on semantic over acoustic information which is critical for a vocoder.

\begin{table}[!h]
 \centering
 \caption{Benchmarking results for the speech synthesis task across various self-supervised learning (SSL) models, both in base and large configurations, using different cluster sizes (k=128, 512, 1024) for speech unit extraction. The Mean Opinion Score (MOS) is used as the performance metric.}
 \label{tab:ss}
 \resizebox{0.5\textwidth}{!}{
 \begin{tabular}{c|c|cccccc}
    \toprule
    \multirow{2}{*}{SSL Model} & \multirow{2}{*}{Setting} & \multicolumn{6}{c}{Layer Base/Large - MOS $\uparrow$} \\
     & & 2/4 & 4/8 & 6/12 & 8/16 & 10/20 & 12/24 \\
     \midrule
     \multirow{3}{*}{Hubert Base} & k=128 & 3.24 & 3.43 & 3.15 & 2.88 & 3.25 & 3.35 \\
     & k=512 & 3.33 & 3.32 & \textbf{3.45} & 3.18 & 3.05 & 3.28 \\
     & k=1024 & 3.33 & 3.33 & 3.26 & 3.13 & 3.17 & 3.32 \\
     \midrule
     \multirow{3}{*}{Wav2Vec2 Base} & k=128 & 3.30 & 3.01 & 2.87 & 2.94 & 3.04 & 3.06 \\
     & k=512 & 3.35 & 3.26 & 3.17 & 3.05 & 3.13 & 2.96 \\
     & k=1024 & 3.41 & 3.26 & 2.90 & 3.17 & 2.79 & 2.79  \\
     \midrule
     \midrule
     \multirow{3}{*}{mHuBERT Base} & k=128 & 3.26 & 3.27 & 3.17 & 3.20 & 2.94 & 3.00 \\
     & k=512 & 3.55 & 3.38 & 3.34 & 3.27 & 3.24 & 3.12 \\
     & k=1024 & 3.47 & 3.40 & 3.41 & 3.24 & 3.25 & 3.25 \\
     \midrule
     \multirow{3}{*}{Wav2Vec2 XLS-R Large} & k=128 & 3.67 & 3.52 & 3.39 & 3.25 & 3.55 & 2.42 \\
     & k=512 & 3.65 & 3.62 & 3.59 & 3.62 & \textbf{3.80} & 2.93 \\
     & k=1024 & 3.69 & 3.53 & 3.51 & 3.50 & 3.69 & 3.14 \\
     \midrule
     \multirow{3}{*}{SAMU-XLSR Large} & k=128 & 3.38 & 3.27 & 2.71 & 2.19 & 1.71 & 1.34 \\
     & k=512 & 3.28 & 3.06 & 3.19 & 2.75 & 2.12 & 1.25 \\
     & k=1024 & 3.47 & 3.21 & 2.80 & 2.80 & 2.30 & 1.57 \\
    \bottomrule
 \end{tabular}
 }
\end{table}

\subsection{Speech to Speech Translation}
In the following section, we evaluate a subset of previous configurations on the speech-to-speech translation task. 
To ensure comparable results with previous studies, we chose to include the HuBERT Base at layer~$6$ with a cluster size of $128$ as the baseline.  
Additionally, to assess the effect of the number of clusters on the translation task, we include results at layer~$6$ with cluster sizes of $512$ and $1024$.  
To understand the impact between semantic and acoustic on the S2ST task, we selected both Wav2Vec2 XLS-R and SAMU-XLSR, choosing configurations that maximize performance on the ASR task (layer~$16$) and configurations that maximize performance on the speech synthesis task (layer~$8$).  \\

\begin{figure}[!t]
    \centering
    \includegraphics[width=0.40\textwidth]{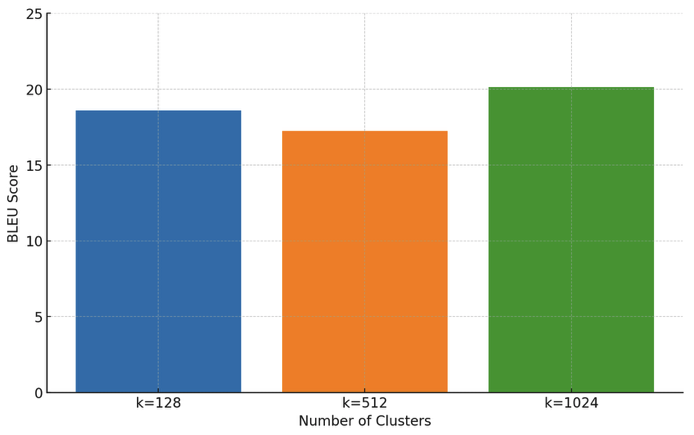}
\caption{Effect of number of clusters BLEU scores. We adopt the baseline configuration, HuBERT Base at layer~$6$ with (k=128, 512, 1024)}
\label{fig:kmeans}
\end{figure}

From Figure~\ref{fig:kmeans}, we can observe that BLEU scores improve as the cluster count increases from $128$ to $1024$, with a minor reduction when transitioning from $128$ to $512$ clusters. It is interesting considering that the configuration Hubert Base with layer~$6$ and $512$ clusters achieved the highest Mean Opinion Score (MOS).
This pattern indicates that a higher number of clusters generally enhances model performance.  
The most effective configuration tested is the Hubert Base model with 1024 clusters, which achieved the highest BLEU score of $20.14$. \\

Looking at the BLEU scores presented in Table~\ref{tab:s2st}, the Wav2Vec2 XLS-R Large model consistently outperforms the SAMU-XLSR Large model across both layer configurations.  
This suggests that the fine-tuning of SAMU-XLSR Large on text embeddings may not effectively contribute to speech-to-speech translation tasks.  
Furthermore, the results on Wav2Vec2 XLS-R Large underscore the complexity in selecting the optimal layer for extracting discrete speech units. Relying solely on the Mean Opinion Score (MOS) for token selection does not appear to yield the best results, compared to adopting a balanced approach that considers both the Character Error Rate (CER) and MOS scores. \\

\begin{table}[!h]
 \centering
 \caption{This table compares the BLEU scores of Wav2Vec2 XLS-R Large and SAMU-XLSR Large models under two distinct configurations: one optimizing for CER performance (layer 8) and the other for maximizing MOS scores (layer 16), both using 1024 clusters.}
 \label{tab:s2st}
 \resizebox{0.5\textwidth}{!}{
 \begin{tabular}{c|c|c|c}
    \toprule
    SSL Model & Layer & Number Of Clusters & BLEU $\uparrow$ \\
    \midrule
    Wav2Vec2 XLS-R  Large & 8 & 1024 & 17.55 \\
    \midrule
    Wav2Vec2 XLS-R  Large & 16 & 1024 & 16.93 \\
    \midrule
    SAMU-XLSR Large & 8 & 1024 & 16.9 \\
    \midrule
    SAMU-XLSR Large & 16 & 1024 & 13.29 \\
    \bottomrule
 \end{tabular}
 }
\end{table}
\section{Conclusion}
In this work, we have described various experiments to evaluate the robustness of discrete speech units in downstream tasks.  
The results obtained in textless speech-to-speech translation underscore the complexity in selecting encoders and the number of clusters.  
We hope this analysis will help the community in better understanding the extraction of discrete tokens.  
Looking forward, future research will explore the combination of multiple encoders and layers within the same task and extend the approach to additional language pairs, particularly those that are unwritten.  
\newpage

\section{Acknowledgements}
This work received funding from the European SELMA project (grant N°957017).

\bibliographystyle{IEEEtran}
\bibliography{mybib}

\end{document}